# EXPLORING A MULTI-STAGE FEEDBACK TEACHING MODE FOR GRADUATE STUDENTS OF SOFTWARE ENGINEERING DISCIPLINE BASED ON PROJECT-DRIVEN COMPETITION

Xiangdong Pei[1,2,3]*, Rui Zhang[2,3]*

[1]College of Computer, National University of Defense Technology, Changsha, China
[2]Shanxi Supercomputing Center, Digital Economic Park of Luliang Economic Development Zone, Tianjiahui Street, Lvliang, China
[3]College of Computer Science and Technology, Taiyuan University of Technology, No. 66, Waliu Road, Wanbailin District, Taiyuan, China
* pxd@nudt.edu.cn

## Abstract

Aiming at the current problems of theory-oriented, practice-light, and lack of innovation ability in the teaching of postgraduate software engineering courses, a multi-stage feedback teaching mode for software engineering postgraduates based on competition project-driven is proposed. The model is driven by the competition project, and implementing suggestions are given in terms of stage allocation of software engineering course tasks and ability cultivation, competition case design and process evaluation improvement, etc. Through the implementation of this teaching mode, students' enthusiasm and initiative are expected to be stimulated, and the overall development of students' professional skills and comprehension ability would be improved to meet the demand of society for software engineering technical talents.

## 1 Introduction

The construction of a new engineering discipline with the direction of cultivating new engineering talents has put forward clear and explicit requirements for the teaching of software engineering courses. As an engineering discipline with strong practicality, it is urgent to solve the crucial problem of cultivating the innovation ability of talents in a software engineering discipline. Strengthening the professional practice of master's students in software engineering disciplines is the primary way to improve practical ability and appreciate innovation awareness. Due to various reasons, the practical innovation ability of Chinese master students in software engineering disciplines is generally not strong. As software engineering technology is developing and updating swiftly, system ability, practical ability, innovation ability, and independent learning ability are especially important [1-2]. At present, full-time graduate students in China predominantly come from fresh undergraduates. Although they have relatively solid theoretical knowledge, they are deficient in poor practical ability and weak innovation consciousness because they do not have any work experience, thus their professional practical innovation ability needs to be further strengthened.

At present, the scholarly teaching mode is commonly adopted in numerous universities to cultivate professional graduate students, most of which pay attention to theoretical teaching with fewer projects combined with production and practical life, neglecting the cultivation of practical innovation ability of master students. In terms of the assessments, only the assessment of theoretical knowledge is emphasized, and the assessment of students' practical innovation ability is not highlighted, which finally forms a phenomenon of "emphasizing theory but not practice". For most students, they don't know what they are good at and what they are interested in, so they don't intentionally reinforce what they are good at. Nevertheless, in software engineering, a person with expertise in one area is better than a person with expertise in several areas. The purpose of the examination is to find out the students' ability and interest in each module, so as to achieve the purpose of personalized training for students in the later stage. At the same time, most of the existing binary tutor systems are limited to the surface level, and the enterprise tutors are not highly motivated and neglected in the guiding process. After the topic is selected, learners can enhance their special ability by completing specific tasks related to the topic.

According to the current situation of cultivating practical innovation ability of master students in software engineering, it is urgent to explore a new and viable method of cultivating practical innovation ability of master students. In this paper, through researching the components, influencing factors, cultivation methods, and discipline characteristics of the practical innovation ability of master students in software engineering disciplines, a cultivation system in line with the development goals of local universities is provided to enhance the practical innovation ability of master students in software engineering disciplines. It is expected that the present research would be helpful to cultivate more and better practical innovative talents who are application-oriented, fully developed, and of superior level for Chinese enterprises and institutions.

## 2. Methodology

*2.1 Multi-stage feedback teaching*

The so-called multi-stage feedback teaching method divides the entire learning process into basic learning, ability enhancement, comprehensive application, process assessment, and feedback evolution. Among them, basic learning, ability enhancement, and comprehensive application are the infrastructure stages, that is, the basic learning stage, ability enhancement stage, and comprehensive application stage (including practical examination and defense and summary) are divided according to the ratio of 4:3:3 according to the total credit hours of the course, and students are encouraged to participate in various enterprise software projects and design competitions in the comprehensive application stage.

One of the basic learning phases is to let students understand various current software design methods and how to measure the complexity of software tasks, to be able to analyze software requirements using methods such as modularity and to express the software design process using standard diagrams, tables, and text of software engineering [3-4]. After the completion of this phase of the study, a process assessment will be conducted on the student's current knowledge, and the focus of each part of the examination questions corresponds to the index points of their developed competencies. Based on the student's performance scores in each part of the test, the students will be initially classified into five directions: business personnel, designers, developers, testers, and managers, after which the competence enhancement phase will continue to focus on the development of Students will continue to focus on developing their professional skills during the competency enhancement phase.

The competency enhancement phase focuses on mastering the application knowledge of the material, such as implementation, maintenance, object-oriented implementation, software project management, and testing. Students' test scores and interests in the above five directions will be combined to form a reasonable group division. Each group should contain five types of people: business people, designers, developers, testers, and managers.

In the comprehensive application stage, comprehensive experimental design and project defense will be arranged. We will improve the teaching of various projects, competitions, and experimental practice, introduce the idea of competition and project-driven training, and select practical projects or simulated competition projects of moderate scale, realistic, typical and innovative as the topics, so that the graduate students can feel the relevance of the professional skills and the actual market career positions in the future, and spark the enthusiasm and initiative of the graduate students. After the comprehensive experimental design is completed, the defense will be organized to find out the difficulties encountered in the development process and to summarize the experience through the defense.

*2.2 Project-driven, competition-promoting learning*

In order to further enhance students' independent learning and innovation consciousness, and to apply the learned knowledge in solving practical problems, the project teaching mode of "project-driven and competition-promoted learning" is introduced based on the multi-stage teaching method.

The notion of the project-driven teaching method [5-6] is to place the entire teaching process in a real project, select a moderate scale and innovative project in the student's learning activities, and then incorporate the knowledge points related to software engineering into each stage of project development. The project-driven teaching method creates classroom teaching with a close connection between theory and practice with the help of the school-enterprise collaboration platform and hires experienced project managers, software engineers, and test engineers from enterprises to teach students. In the teaching design of software engineering course, the actual projects of enterprises with moderate scale, realistic, typical and innovative are selected as topics, and students are required to divide into project teams, and simulate the roles in the actual projects The project is based on the project, the teacher is the guide, and the students perform their main role to complete the project with the theoretical foundation and development skills they have mastered.

The competition-driven teaching method [6-8] is a method in which the teacher encourages, organizes, and guides students to participate in a highly targeted and operational skills competition. The advantageous feature of the competition-driven teaching method is that the competition can convert theoretical knowledge into competition content, learn knowledge in competition, identify gaps in competition, and achieve the purpose of enhancing students' comprehensive skills. The school promotes and organizes students to participate in a standardized competition held by the Ministry of Education, a company, or an institution. The educator comprehensively analyzes the requirements of the competition theme, competition protocol, and competition assessment standards integrate the competition examination points into the course teaching, and evaluate students' practical ability and innovation ability by adopting the assessment form of competition mechanism and competitive situation, so as to stimulate students' active learning. Since the competition covers a wide range of knowledge points, its assessment standards not only correspond to the key contents of software majors but also closely match the actual cases of enterprise engineering, so the competition-oriented teaching method is an effective method of teaching reform and complements the shortcomings of conventional teaching.

The introduction of the teaching method of "competition + teaching" has increased the proportion of practical hands-on ability and programming content in the classroom, which stimulates new ideas of problem-solving and develops students' innovative thinking in the process of doing it by themselves and adapts to the demand for innovative talents in the computer field.

*2.3 Process evaluation assessment*

Software engineering project teaching should be fully combined with the characteristics of the course, abandoning the single test paper form of assessment, switching to an assessment method that focuses on students' learning process and comprehensive ability [9], and forming a diversified assessment mechanism that encompasses classroom assessment, lab report, project completion, participation in competitions and other contents. After students complete classroom learning, classroom quizzes, mutual evaluation among project groups, mutual evaluation within groups, final project defense, competition performance and other assessment means are used to focus on guiding students to change their learning style, strengthening the application of assessment means, permitting students to check the gaps in the comprehensive assessment, mainly solving project difficulties and enhancing professional skills; meanwhile, master's degree students are actively organized to participate in various competitions to exercise students' teamwork ability, widen At the same time, we actively organize master's degree students to participate in various competitions to exercise students' teamwork ability, expand their knowledge and broaden their horizons, and include the project-related patents, papers and science and technology competition awards in the assessment, so as to eventually achieve the effect of "promoting learning through competitions".

*2.4 Feedback evolution*

By splitting the entire learning process of students into three stages, basic learning, ability improvement, comprehensive application, and process assessment, so as to achieve the effect of step-by-step progress in a progressive manner. After the completion of the basic learning stage, a test will be conducted to classify students' interest points in learning, and the problems that appear in the basic learning stage will be recorded; in the ability improvement stage, a good learning plan will be formulated for students according to their interest points in the essential learning, and the deficiencies in the learning process will be recorded; In the extensive application stage, students will be arranged according to their previous learning situation to participate in various competitions. In the comprehensive application stage, according to the previous learning of the students, they are arranged to participate in various competitions to practice what they have learned in the competition. At the same time, the problems in the comprehensive application stage and the feedback from the preceding two stages are summarized and analyzed to guide the teaching in the subsequent academic year.

In summary, the multi-stage feedback teaching mode of software engineering graduate students based on the competition project is constructed as shown in Fig. 1, with the competition project as the driver, and the implementation method is provided in terms of the stage allocation of software engineering course tasks and ability cultivation, the design of the competition case, and the improvement of the process evaluation. At the same time, based on the integration of production and education, it promotes the integration of software engineering professional degree graduate training and industry needs. The actual teaching results illustrate that the integration mode of production and education plays a great role in upgrading students' software engineering ability and satisfying the demand for software development talents of enterprises.

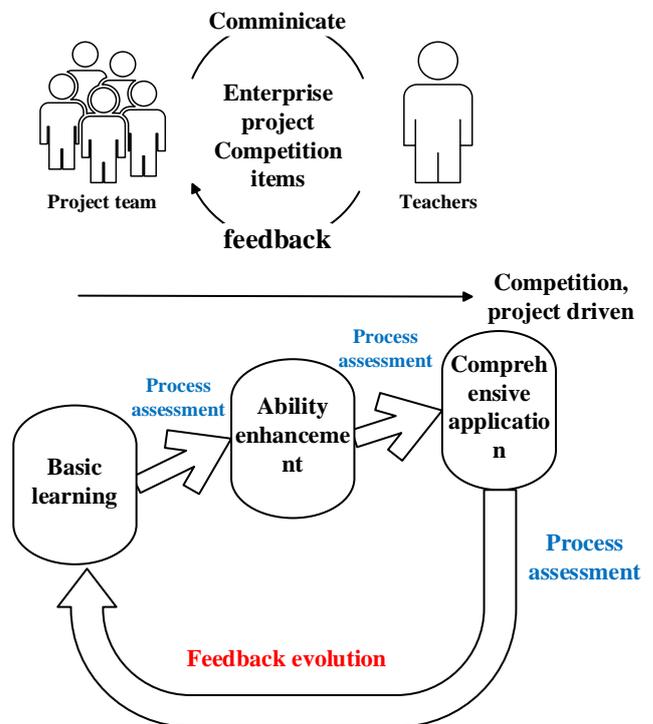

Fig. 1 Multi-stage feedback teaching model for software engineering graduate students based on project-driven competition

## 3 Effectiveness of practical teaching

As a tutor of software engineering graduate students at Taiyuan University of Science and Technology, the author conducts multi-stage feedback teaching for three years of graduate students of software engineering at Taiyuan University of Science and Technology from 19 to 21, focusing on the progression of knowledge points in the teaching process and providing comprehensive guidance to students. (1) Students seriously change from passive knowledge to active knowledge in the classroom, particularly when enterprises come into the classroom, which increases the initiative of students to ask questions and actively participate in the course project, so that theory and practice are closely merged to recognize and learn; (2) The team in the laboratory class starts to have a clear division of labor and supervise each other, and the team starts to actively communicate with each other and within the team to resolve questions, so the absenteeism situation is substantially diminished (3) The proportion of students participating in school-enterprise collaboration projects and competitions has actively participating in this major has reached 60%, and the results of course learning have been transformed and obtained several software copyrights and won a number of awards in competitions at all levels.

## 4 Summary

The cultivation of the practical innovation ability of master's students in software engineering is an important research topic in graduate education. The proposed multi-stage feedback teaching mode for software engineering graduate students based on competition project drive organically combines the stages of basic knowledge, ability enhancement, comprehensive application, process assessment, and feedback evolution, and inserts the competition project drive method into the actual teaching design session to comprehensively strengthen the practical innovation ability of master students in software engineering disciplines. The teaching effect illustrates that the teaching mode effectively promotes students' thinking ability, creativity, and collaborative spirit, which has guiding significance for the improvement of teaching quality in the new stage and provides an effective way for the cultivation of new engineering talents.

## 5 Acknowledgements

This work is supported by the Exploration and evaluation of the cooperative cultivation mode of intelligent software engineering graduate school and enterprises in the context of new engineering (2021YJJG244) and the Exploration and practice of the talent cultivation mode of intelligent science and technology in the context of new engineering (J2021429) of the Teaching Reform Innovation Project of Higher Education Institutions in Shanxi Province.

## 6 References

[1] Xiang Ying, D., Rong, B., Dai Hong J.: 'Research on the cultivation mode of applied talents in software engineering in local universities', China Educational Technology Equipment, 2016, (12), pp 62-64.
[2] Yoga, L.: 'Application of Software Engineering Technology in System Software Development ', Computer Products and Distribution, 2019,7, pp25-29
[3] Qiang, L.: 'Software Engineering Curriculum Reform and practice in local universities based on MOOC/SPOC ', Computer education,2018,2, pp38-42.
[4] Garousi, V., Giray, G., Tuzun, E., et al. 'Closing the gap between software engineering education and industrial needs', IEEE Software,2020,37,（2）, pp68-77.
[5] Li Nan, G.: 'Exploration on the Teaching Mode of "Project-driven and Learning Promoted by Competition" in Colleges and Universities', Journal of PanZhiHua University, 2010.2017, 34, (1), pp 58-63.
[6] Pei Yun, Z.: 'Research on Project-driven Software Engineering Course Case-segmented Situational Teaching', Computer Education,2013, 21, (3), pp77-79
[7] Xu Yang, F., Ri, X., Xiao Kun, Z., et al.: 'Reform and Exploration of College Liberal Arts Computer Basic Course Based on Task, Project and Competition', Educational Theory and Practice, 2017, 37,36, pp 49-51
[8] Cheng, Z., Ping, G., et al.: 'Exploration of Competition project-driven Software Engineering Course Teaching Reform', Computer Education,2018, ,8, pp 22-24
[9] Gang, L., Tao, G., et al.: 'A Brief analysis of the Reform of process-based Comprehensive Curriculum Design Assessment and Evaluation Method', Education and Teaching Forum,2021, 28, pp68-71